\documentclass[aps,prl,preprint,tightenlines,superscriptaddress,
showpacs,byrevtex,floatfix]{revtex4}
\usepackage{epsfig}% Include figure files

\newcommand{\mb}{{M_{\rm bc}}}
\newcommand{\de}{{\Delta{E}}}
\newcommand{\pL}{{p\bar{\Lambda}}}
\newcommand{\plpi}{{p\bar{\Lambda}\pi^-}}

\newcommand{\bp}{{B^{+}}}
\newcommand{\bz}{{B^{0}}}

\newcommand{\ppk}{{p\bar{p}K^+}}

\newcommand{\plpiz}{{p\bar{\Lambda}\pi^0}}
\newcommand{\plg}{{p\bar{\Lambda}\gamma}}

\begin{document}
 
\preprint{\vbox{ \hbox{   }
                 \hbox{\bf Belle Preprint 2007-19}
                 \hbox{\bf KEK Preprint 2007-6}
               % \hbox{ICHEP2006-xx}
%                 \hbox{hep-ex nnnn, if available}
}}
%@twocolumnfalse\endcsname
\title{ \quad\\[0.5cm] Study of $\bp \to \plg$, $\plpiz$ and $\bz \to \plpi$}
%\author{ Yen-Jie Lee and Min-Zu Wang for the Belle Collaboration}
  
%\address{
%Department of Physics, National Taiwan University, Taipei, Taiwan,
%R.O.C.}
%%% Paper:    B -> p Lambda gamma/pi
%%% Journal:  Physical Review D
%%% Contacts: Y.-J. Lee (listeve@hep1.phys.ntu.edu.tw)
%%%           M. Wang (mwang@phys.ntu.edu.tw)
%%% Non-responding authors or those who said NO are commented out.
%%% ====================================================================
%%% Click the RELOAD button on your web browser to see the updated file.
%%% ====================================================================
%%% Use \input{author} to insert this material into your latex file.
%%%%% Force institutions to appear in alphabetical order when typeset.
\affiliation{Budker Institute of Nuclear Physics, Novosibirsk}
%%%\affiliation{Chiba University, Chiba}
\affiliation{University of Cincinnati, Cincinnati, Ohio 45221}
\affiliation{Department of Physics, Fu Jen Catholic University, Taipei}
%%%\affiliation{Justus-Liebig-Universit\"at Gie\ss{}en, Gie\ss{}en}
\affiliation{The Graduate University for Advanced Studies, Hayama}
%%%\affiliation{Gyeongsang National University, Chinju}
\affiliation{Hanyang University, Seoul}
\affiliation{University of Hawaii, Honolulu, Hawaii 96822}
\affiliation{High Energy Accelerator Research Organization (KEK), Tsukuba}
%%%\affiliation{Hiroshima Institute of Technology, Hiroshima}
%%%\affiliation{University of Illinois at Urbana-Champaign, Urbana, Illinois 61801}
\affiliation{Institute of High Energy Physics, Chinese Academy of Sciences, Beijing}
\affiliation{Institute of High Energy Physics, Vienna}
\affiliation{Institute of High Energy Physics, Protvino}
\affiliation{Institute for Theoretical and Experimental Physics, Moscow}
\affiliation{J. Stefan Institute, Ljubljana}
\affiliation{Kanagawa University, Yokohama}
\affiliation{Korea University, Seoul}
%%%\affiliation{Kyoto University, Kyoto}
\affiliation{Kyungpook National University, Taegu}
\affiliation{Swiss Federal Institute of Technology of Lausanne, EPFL, Lausanne}
\affiliation{University of Ljubljana, Ljubljana}
\affiliation{University of Maribor, Maribor}
\affiliation{University of Melbourne, School of Physics, Victoria 3010}
\affiliation{Nagoya University, Nagoya}
\affiliation{Nara Women's University, Nara}
\affiliation{National Central University, Chung-li}
\affiliation{National United University, Miao Li}
\affiliation{Department of Physics, National Taiwan University, Taipei}
\affiliation{H. Niewodniczanski Institute of Nuclear Physics, Krakow}
\affiliation{Nippon Dental University, Niigata}
\affiliation{Niigata University, Niigata}
%%%\affiliation{University of Nova Gorica, Nova Gorica}
\affiliation{Osaka City University, Osaka}
\affiliation{Osaka University, Osaka}
\affiliation{Panjab University, Chandigarh}
\affiliation{Peking University, Beijing}
%%%\affiliation{University of Pittsburgh, Pittsburgh, Pennsylvania 15260}
%%%\affiliation{Princeton University, Princeton, New Jersey 08544}
\affiliation{RIKEN BNL Research Center, Upton, New York 11973}
%%%\affiliation{Saga University, Saga}
\affiliation{University of Science and Technology of China, Hefei}
\affiliation{Seoul National University, Seoul}
\affiliation{Shinshu University, Nagano}
\affiliation{Sungkyunkwan University, Suwon}
\affiliation{University of Sydney, Sydney, New South Wales}
\affiliation{Tata Institute of Fundamental Research, Mumbai}
\affiliation{Toho University, Funabashi}
\affiliation{Tohoku Gakuin University, Tagajo}
\affiliation{Tohoku University, Sendai}
\affiliation{Department of Physics, University of Tokyo, Tokyo}
\affiliation{Tokyo Institute of Technology, Tokyo}
\affiliation{Tokyo Metropolitan University, Tokyo}
\affiliation{Tokyo University of Agriculture and Technology, Tokyo}
%%%\affiliation{Toyama National College of Maritime Technology, Toyama}
%%%\affiliation{University of Tsukuba, Tsukuba}
\affiliation{Virginia Polytechnic Institute and State University, Blacksburg, Virginia 24061}
\affiliation{Yonsei University, Seoul}
  \author{M.-Z.~Wang}\affiliation{Department of Physics, National Taiwan University, Taipei} % Taiwan
 \author{Y.-J.~Lee}\affiliation{Department of Physics, National Taiwan University, Taipei} % Taiwan
% \author{K.~Abe}\affiliation{High Energy Accelerator Research Organization (KEK), Tsukuba} % KEK
  \author{K.~Abe}\affiliation{Tohoku Gakuin University, Tagajo} % TohokuGakuin
% \author{N.~Abe}\affiliation{Tokyo Institute of Technology, Tokyo} % TIT
  \author{I.~Adachi}\affiliation{High Energy Accelerator Research Organization (KEK), Tsukuba} % KEK
  \author{H.~Aihara}\affiliation{Department of Physics, University of Tokyo, Tokyo} % Tokyo
  \author{D.~Anipko}\affiliation{Budker Institute of Nuclear Physics, Novosibirsk} % BINP
% \author{K.~Aoki}\affiliation{Nagoya University, Nagoya} % Nagoya
% \author{K.~Arinstein}\affiliation{Budker Institute of Nuclear Physics, Novosibirsk} % BINP
% \author{Y.~Asano}\affiliation{University of Tsukuba, Tsukuba} % Tsukuba
% \author{T.~Aso}\affiliation{Toyama National College of Maritime Technology, Toyama} % Toyama
% \author{V.~Aulchenko}\affiliation{Budker Institute of Nuclear Physics, Novosibirsk} % BINP
  \author{T.~Aushev}\affiliation{Swiss Federal Institute of Technology of Lausanne, EPFL, Lausanne}\affiliation{Institute for Theoretical and Experimental Physics, Moscow} % ITEP
% \author{T.~Aziz}\affiliation{Tata Institute of Fundamental Research, Mumbai} % Tata
  \author{S.~Bahinipati}\affiliation{University of Cincinnati, Cincinnati, Ohio 45221} % Cincinnati
  \author{A.~M.~Bakich}\affiliation{University of Sydney, Sydney, New South Wales} % Sydney
% \author{V.~Balagura}\affiliation{Institute for Theoretical and Experimental Physics, Moscow} % ITEP
% \author{Y.~Ban}\affiliation{Peking University, Beijing} % Peking
% \author{S.~Banerjee}\affiliation{Tata Institute of Fundamental Research, Mumbai} % Tata
% \author{E.~Barberio}\affiliation{University of Melbourne, School of Physics, Victoria 3010} % Melbourne
% \author{M.~Barbero}\affiliation{University of Hawaii, Honolulu, Hawaii 96822} % Hawaii
% \author{A.~Bay}\affiliation{Swiss Federal Institute of Technology of Lausanne, EPFL, Lausanne} % Lausanne
  \author{I.~Bedny}\affiliation{Budker Institute of Nuclear Physics, Novosibirsk} % BINP
% ...
% Ljubljana
% \author{D.~Z\"urcher}\affiliation{Swiss Federal Institute of Technology of Lausanne, EPFL, Lausanne} % Lausanne
\collaboration{The Belle Collaboration}

\normalsize
\begin{abstract}
We study the following charmless baryonic
three-body decays of $B$ mesons:
$\bp \to \plg$, $\bp \to \plpiz$ and $\bz \to \plpi$.
The partial
branching fractions as a function of the baryon-antibaryon mass and
the polar angle distributions of
the proton in the baryon-antibaryon system are presented.
This study includes the first observation of $\bp\to\plpiz$,  
which is measured to have a branching fraction of 
$(3.00^{+0.61}_{-0.53}\pm 0.33) \times 10^{-6}$. 
We also set upper limits on branching fractions of the
two-body decays 
$\bz \to p {\bar\Sigma}^{*-}$, $\bz \to \Delta^0 \bar{\Lambda}$,
$\bp \to p {\bar\Sigma}^{*0}$, and  $\bp \to \Delta^+ \bar{\Lambda}$
%${\mathcal B}(\bp \to p \Delta^0) < 2.3 \times 10^{-7}$,
%${\mathcal B}(\bp \to \bar{p} \Delta^{++) < 2.3 \times 10^{-7}$,
at the 90\% confidence level.
These results are obtained from a $414\,{\rm fb}^{-1}$ data sample
%that contains 449 $\times 10^6 B\bar{B}$ pairs 
collected near the $\Upsilon(4S)$ resonance
with the Belle detector at the KEKB asymmetric-energy $e^+ e^-$
collider.
 
\vskip 1cm
\end{abstract}
 
\maketitle

{\renewcommand{\thefootnote}{\fnsymbol{footnote}}}
\setcounter{footnote}{0}

\begin{figure}[htb]
\begin{center}

\epsfig{file=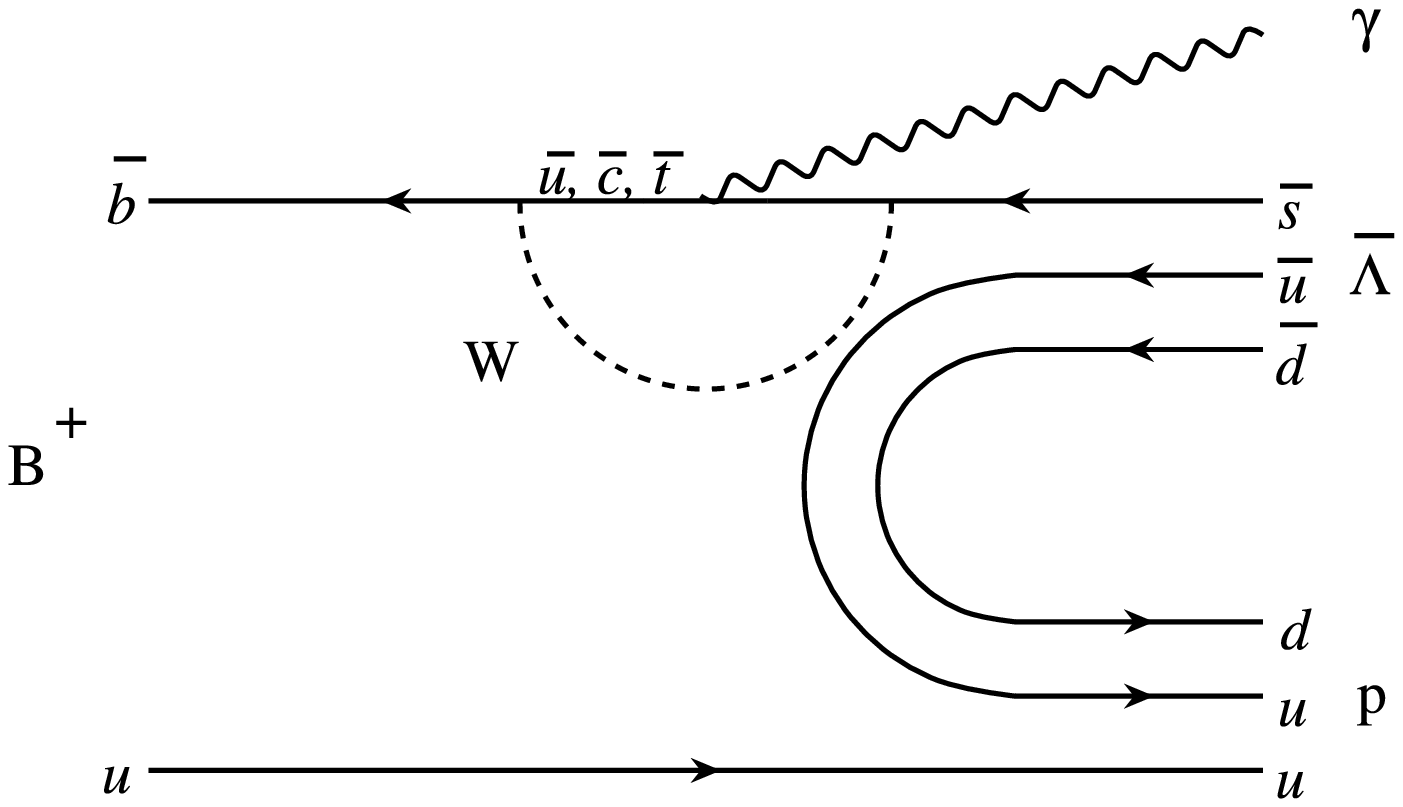, width=4in}
\caption{ 
A $b \to s$ penguin diagram for $\bp \to \plg$. 
}
\label{fg:pLg}
\end{center}
\end{figure}

After the first observation of charmless baryonic $B$ meson decay,
$\bp \to \ppk$~\cite{ppk,conjugate}, various three-body baryonic decays were
found~\cite{plpi,LLK,plg}. The dominant contributions for these decays
are presumably via the $b \to s$ penguin diagram as shown in 
Fig.~\ref{fg:pLg} for the case of $\bp \to \plg$.
A common experimental  
feature of these decays is that the baryon-antibaryon mass 
spectra peak near threshold. This feature was conjectured 
in Ref.~\cite{HS} and has recently aroused much 
theoretical interest~\cite{theory}. 
Detailed information from
the polar angle  distributions~\cite{polar} and Dalitz plot~\cite{BaBarppk}
offer better understanding of the underlying dynamics.

In this paper, we study the following three-body charmless baryonic decays of $B$ mesons: 
$\bp \to \plg$, $\bp \to \plpiz$ 
and $\bz \to \plpi$. 
The partial 
branching fractions as a function of the baryon-antibaryon mass and  
the polar angle distributions of 
the proton in the baryon-antibaryon system are presented.
It is interesting to compare the results
with theoretical predictions~\cite{HandC,gengg}.
Since the $\Lambda$ hyperon could be a useful tool
to probe the helicity selection rule for 
the $b \to s$ process~\cite{HandC,Suzuki}, we investigate  
the proton polar angular distribution from $\Lambda$ decays.  
We also search for  
intermediate two-body decays in these three-body final states. This is 
motivated by the observations of two-body decays of charmed 
baryons~\cite{Lcp}.
Using topological quark diagrams for $B$ decays and the assumption of 
SU(3) flavor symmetry,
various two-body charmless baryonic decay modes
should be observable with a data sample of $\sim$400 fb$^{-1}$~\cite{chua}.

We use a  414 fb$^{-1}$  data sample
consisting of 449 $ \times 10^6 B\bar{B}$ pairs
collected with the Belle detector %on the $\Upsilon({\rm 4S})$ resonance
at the KEKB asymmetric energy $e^+e^-$ (3.5 on 8~GeV) collider~\cite{KEKB}.
The Belle detector is a large-solid-angle magnetic
spectrometer that
consists of a silicon vertex detector (SVD),
a 50-layer central drift chamber (CDC), an array of
aerogel threshold Cherenkov counters (ACC),
a barrel-like arrangement of time-of-flight
scintillation counters (TOF), and an electromagnetic calorimeter (ECL)
composed of CsI(Tl) crystals located inside
a superconducting solenoid coil that provides a 1.5~T
magnetic field.  An iron flux-return located outside
the coil is instrumented to detect $K_L^0$ mesons and to identify
muons.  The detector
is described in detail elsewhere~\cite{Belle}.
The following two kinds of inner detector 
configurations were used. A 2.0 cm beam pipe
and a 3-layer silicon vertex detector was used for the first sample
of 152 $\times 10^6 B\bar{B}$ pairs, while a 1.5 cm beam pipe, a 4-layer
silicon detector and a small-cell inner drift chamber were used to record
the remaining 297 $\times 10^6 B\bar{B}$ pairs~\cite{Ushiroda}.

The event selection criteria are based on the information obtained
from the tracking system
(SVD and CDC) and the particle identification system (CDC, ACC, TOF and ECL).
They are optimized using
Monte Carlo (MC) event samples produced by 
the EvtGen generator~\cite{evtgen} and
GEANT~\cite{geant} detector simulation.
All primary charged tracks
are required to satisfy track quality criteria
based on the track impact parameters relative to the
interaction point (IP). %, which is determined run-by-run.
The deviations from the IP position are required to be within
$\pm$0.3 cm in the transverse ($x$--$y$) plane, and within $\pm$3 cm
in the $z$ direction, where the $+z$ axis is opposite to the
positron beam direction. For each track, the likelihood values $L_p$,
$L_K$, and $L_\pi$ that it is a proton, kaon, or pion, respectively,
are determined from the information provided by
the particle identification system.  The track is identified as a proton
if $L_p/(L_p+L_K)> 0.6 $ and $L_p/(L_p+L_{\pi})> 0.6$, or as a pion 
if $L_{\pi}/(L_K+L_{\pi})> 0.6$.
%based on
%using $p/K/\pi$ likelihood functions obtained from the hadron
%identification system. For protons, we require $L_p/(L_p+L_K)> 0.6 $ and
%$L_p/(L_p+L_{\pi})> 0.6$, where $L_{p/K/\pi}$ stands for the
%proton/kaon/pion likelihood.
For charged particles with
momenta around 2 GeV/$c$,
the proton selection efficiency is about 84\% (88\% for $p$ and 80\% for 
$\bar{p}$ due to larger $\bar{p}$ cross sections)
and the fake rate is about
10\% for kaons and 3\% for pions.
Candidate $\Lambda$ baryons are reconstructed from pairs of oppositely
charged tracks---treated as a proton and negative pion---whose  mass
is consistent with the nominal $\Lambda$ baryon mass, 
1.111 GeV/$c^2 < M_{p\pi^-} <1.121 $ GeV/$c^2$.  
The $\Lambda$ candidate
 should have a displaced vertex and 
its momentum direction should be 
consistent with
a $\Lambda$ originating from the IP position.
For particle identification of the $\Lambda$ daughters (i.e. secondary
charged tracks), we require only
$L_p/(L_p+L_{\pi})> 0.6$ for the proton, but do not impose 
any additional requirements on
$L_p/(L_p+L_K)$ for the proton or
$L_{\pi}/(L_K+L_{\pi})$ for the pion.
%The proton-like
%daughter is required to satisfy $L_p/(L_p+L_{\pi})> 0.6$.
%via the $p\pi^-$ decay channel 
%following similar selection procedure for $\ks$. To reduce background, 
%a $L_p/(L_p+L_{\pi})> 0.6$ requirement is applied 
%to the secondary proton from $\Lambda$ decay. 
Photon candidates are selected from neutral clusters in the ECL.
Pairs of photons with invariant masses in the range 115 MeV/$c^2$  $< 
m_{\gamma\gamma}  < 152 $ 
MeV/$c^2$ are used to form $\pi^0$ mesons. 
The measured energy of each photon in the laboratory frame is required to
be greater than 50 MeV. The momentum of the $\pi^0$ 
in the laboratory frame
should be greater than 200 MeV/c. 
The cosine of the decay angle should satisfy $|\cos\theta_\gamma| < 0.9$, 
where $\theta_\gamma$ is  
%the angle in the $\pi^0$ rest frame 
%between the photon direction and 
%the $\pi^0$ direction in the laboratory frame.
the angle between the photon direction and 
the negative of the laboratory frame direction in the $\pi^0$ rest frame.  
The primary photon from the $\bp \to \plg$ decay must satisfy the following
additional
requirements: it should be in
the barrel region (with polar angle between $33^\circ$ and
$128^\circ$) and have an energy greater than 500 MeV.
We discard the primary photon candidate if, in combination with any
              other photon above 30 (200) MeV, its mass 
              is within $\pm 18$ ($\pm
              32$) MeV/$c^2$ of the nominal mass of the $\pi^0$ ($\eta$)
              meson.  
%The above selection criteria are optimized using
%Monte Carlo (MC) simulated event samples.

%Since the initial collision 
%center of mass energy is set to match the $\Upsilon({\rm
%4S})$ resonance, which decays into a $B\bar{B}$ pair, one can use
Candidate $B$ mesons are reconstructed in the 
$\bp \to \plg$, $\bp \to \plpiz$ 
and $\bz \to \plpi$ modes.
We use two kinematic variables in the center-of-mass (CM) frame to identify the
reconstructed $B$ meson candidates: the beam energy
constrained mass $\mb = \sqrt{E^2_{\rm beam}-p^2_B}$, and the
energy difference $\de = E_B - E_{\rm beam}$, where $E_{\rm
beam}$ is the beam energy, and $p_B$ and $E_B$ are the momentum and
energy, respectively, of the reconstructed $B$ meson.
%, all in the rest frame of
%the $\Upsilon({\rm 4S})$.k
The candidate region is
defined as 5.20 GeV/$c^2 < \mb < 5.29$ GeV/$c^2$ and $-0.16$ GeV $ < \de< 0.5$
GeV for the $\pi^0/\gamma$ mode ($-0.1$ GeV $ < \de< 0.3$ GeV for the $\pi^-$ 
mode). 
%From a GEANT~\cite{geant} based Monte Carlo (MC) simulation, 
The signal
peaks in the subregion 5.27 GeV/$c^2 < \mb < 5.29$ GeV/$c^2$ and  
$-0.135$ GeV$<\de<0.074$ GeV for the $\pi^0/\gamma$ mode
($|\de|< 0.03$ GeV for the $\pi^-$
mode). 
The lower bound of $\de$ is chosen to exclude possible contamination from
so-called ``cross-feed'' baryonic $B$ decays, i.e. four-body decays 
with a missed daughter.
%feeding into three-body modes. 

The background in the candidate region 
arises predominantly from the $e^+e^-
\to q\bar{q}$ ($q = u,\ d,\ s,\ c$) continuum.
% Owing to the $\de > $ -0.1 GeV  selection, 
%the  contributions from ``cross-feed'', where similar 
%types of rare decay events pass each other's signal criteria, is negligible.
%Except for a small 
%feed-across between these rare decay modes,
%the dominant background
%is from the continuum $e^+e^- \to q\bar{q}$ process.
%The background from $b \to c$ and charmless mesonic decays is 
%also negligible.
%This is confirmed using
%an off-resonance data set (8.8 fb$^{-1}$)
%taken 60 MeV
%below the $\Upsilon({\rm 4S})$ and a MC sample of 60 million continuum events.
We suppress the jet-like continuum background relative to the more
spherical $B\bar{B}$ signal using a Fisher discriminant~\cite{fisher}
that combines seven event shape variables as described in Ref.~\cite{etapk}.
The $\bp \to \plpiz$ mode has more background than the other modes
and therefore we add the 
missing mass to the Fisher variable. The missing mass is determined
from the rest of the detected particles (treated as 
charged pions or photons) in the event
% assuming they
%are pions for charged particles and photons for neutral particles.  
assuming they 
are decay products 
of the other $B$ meson.
%In the $\Upsilon({\rm 4S})$ rest frame,
%continuum events are jet-like while
%$B\bar{B}$ events are more spherical. 
%One can use the reconstructed momenta 
%of final state particles to form various shape variables (e.g. thrust
%angle, Fox-Wolfram moments, etc.) in order to categorize each event.  
%We follow the scheme defined in Ref.~\cite{etapk} that
%combines seven event shape variables into
%a Fisher discriminant~\cite{fisher} in order to suppress
%continuum background. 
%The variables chosen have
%almost no correlation
%with $\mb$ and $\de$.
We form the signal (background)
likelihood ${\mathcal L}_{s}$ (${\mathcal L}_{b}$) by combining
probability density functions (PDFs) for the Fisher discriminant and
the cosine of the angle between the $B$ flight direction
and the beam direction in the $\Upsilon({\rm 4S})$ rest frame. 
The signal PDFs are determined using signal MC
simulation; the background PDFs are obtained from 
the side-band data %the continuum MC
%simulation for events with
with $\mb < 5.26$ GeV/$c^2$.
We require
the likelihood ratio ${\mathcal R} 
= {\mathcal L}_s/({\mathcal L}_s+{\mathcal L}_b)$ 
to be greater than 0.75, 0.85, and 0.80 for the
$\plg$, $\plpiz$ and $\plpi$ modes, respectively.
These selection
criteria are determined by optimizing $n_s/\sqrt{n_s+n_b}$, where $n_s$ 
and $n_b$
denote the expected numbers of signal and background events, respectively. 
%Note that a nominal signal branching fraction~\cite{ppk,plpi,pph} is assumed
%for each mode to determine $N_s$.
We use the branching fractions from our 
previous measurements~\cite{plg,polar} 
in the calculation of $n_s$. The branching fraction
of $\bp \to \plpiz$ is assumed to be one half that for
$\bz \to \plpi$~\cite{HandC}.
%suppressing about 94\% of the background while retaining 66\% of the signal.
%In this study, there is only one $B$ candidate allowed per event. 
If there are  multiple $B$ candidates in a single event, we 
select the one with the best ${\mathcal R}$ value. 
%Note that there is a special $\Lambda_c^+$ veto for the $\bz \to \plpi$ mode.
We apply a $\Lambda_c^+ \to \Lambda\pi^+$ veto for the
$\bz \to \plpi$ mode: candidate events with a 
reconstructed $\Lambda\pi^+$ mass
in the range 2.26-2.31 GeV/$c^2$ are excluded.%from the vertex fit.
% (including the $K_s/|Lambda$ secondary vertex if applicable).
%in which
%only the primary charged tracks %and IP information 
%are used.

We perform an unbinned extended likelihood fit 
that maximizes the likelihood function, 
%Events are fitted using the extended likelihood function:
$$ L = {e^{-(N_s+N_b)} \over N!}\prod_{i=1}^{N} 
\left[\mathstrut^{\mathstrut}_{\mathstrut}N_s P_s(M_{{\rm bc}_i},\Delta{E}_i)+
N_b P_b(M_{{\rm bc}_i},\Delta{E}_i)\right],$$
to estimate the signal yield in the candidate region.
%5.20 GeV/$c^2 < \mb < 5.29$ GeV/$c^2$ and $-0.1$ GeV $ < \de< 0.2$
%GeV;
Here $P_s\ (P_b)$ denotes the signal (background) PDF, 
$N$ is the number of events in the fit, $i$ is the event index, 
and $N_s$ and $N_b$
are fit parameters representing the number of signal and background
events, respectively.

%The PDFs
%are a Gaussian function to represent the signal $\mb$
%and a double Gaussian for $\de$ with parameters determined by
%MC simulation.
For the signal PDF, we use
two-dimensional functions approximated by smooth histograms
obtained from MC simulation.
%we use the product of a Gaussian in $\mb$ and a double Gaussian in $\de$.
%We fix
%the parameters of these functions to values determined by MC simulation
%~\cite{correction}.
The continuum background PDF 
%assumed to be a product function of uncorrelated $\mb$ and $\de$ shapes.
is taken as the product of shapes in
$\mb$ and $\de$, which are assumed to be uncorrelated.
%These shapes are obtained from sideband
%events, with 0.1 GeV $ < \de < 0.2$ GeV for the $\mb$ function and
%with 5.20 GeV/$c^2$ $ < \mb <$ 5.26 GeV/$c^2$ for the $\de$ function.
%They have been cross checked against a continuum MC sample.
We use an ARGUS~\cite{Argus} parameterization, 
$ f(\mb)\propto \mb\sqrt{1-x^2}
\exp[-\xi (1-x^2)]$,  
%background parametrization first used by the ARGUS collaboration~\cite{Argus} 
to model
the $\mb$ background, with $x$ given by $\mb/E_{\rm beam}$ and $\xi$ as
a fit parameter. %Note that $\mb$ is required to be smaller than $\de$. 
The $\de$ background shape is modeled by a normalized  
second-order polynomial  
whose coefficients are fit parameters.

\begin{figure}[htb]
\begin{center}

\epsfig{file=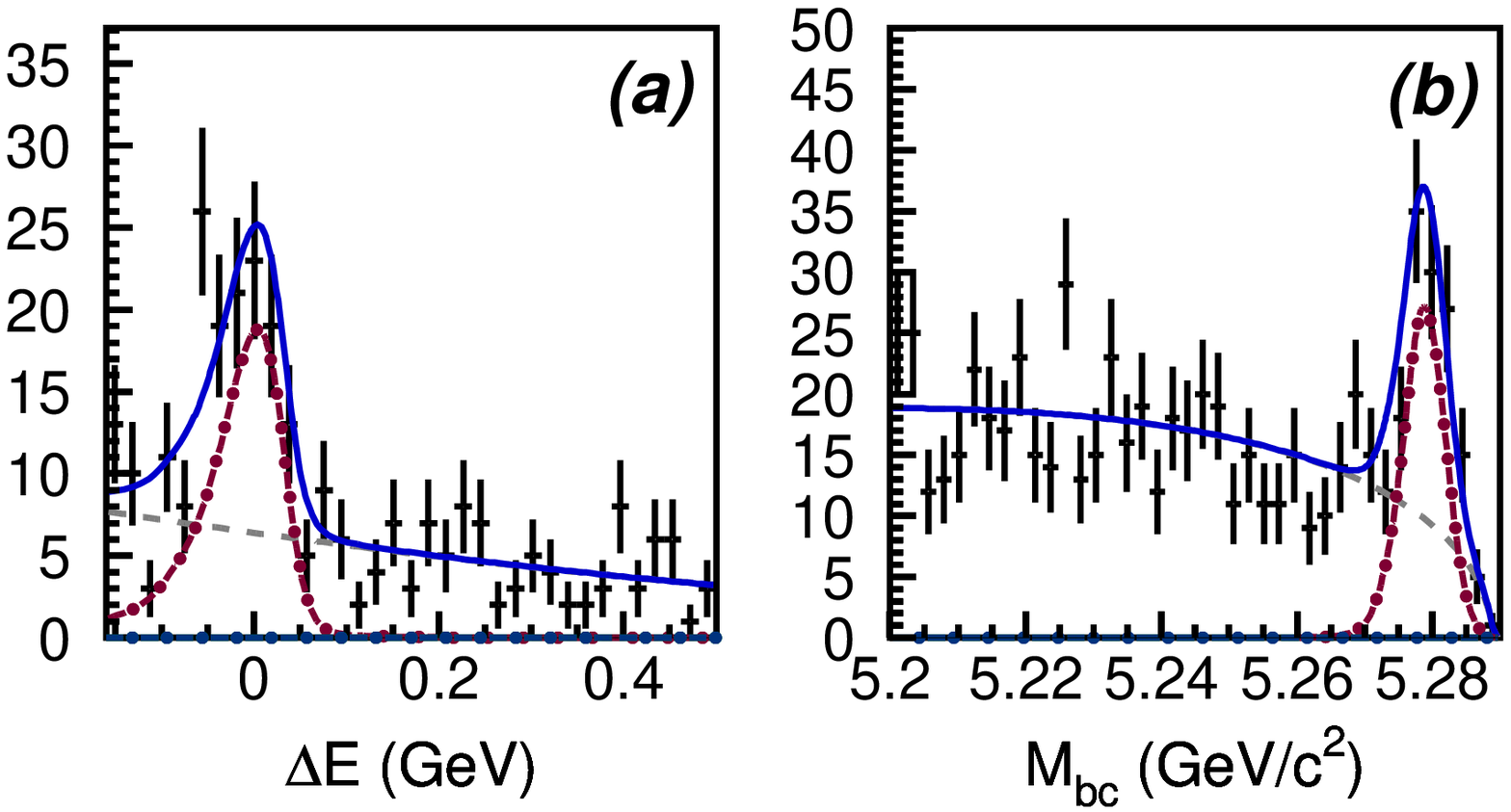, width=3.5in}
\epsfig{file=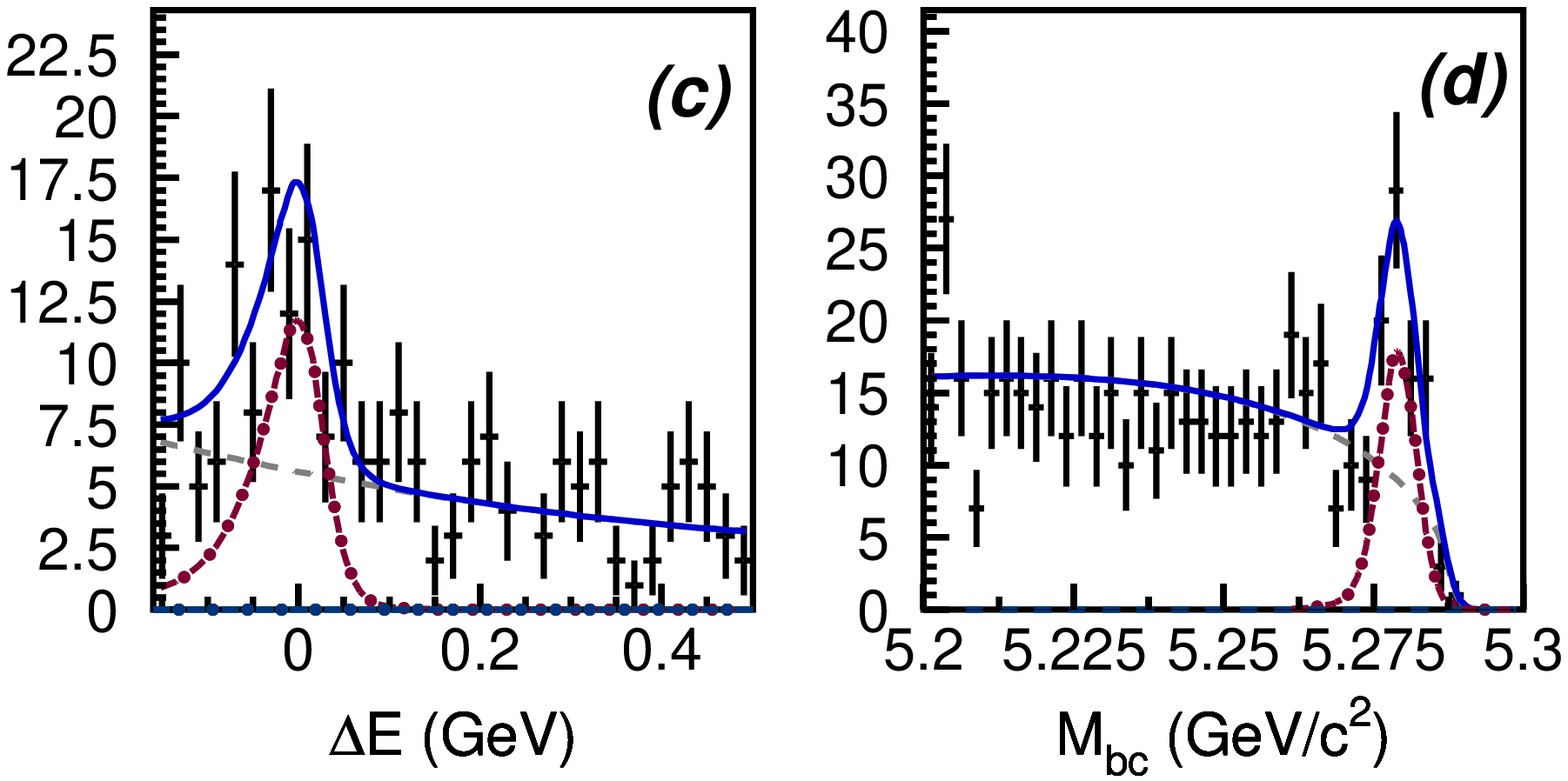, width=3.5in}
\epsfig{file=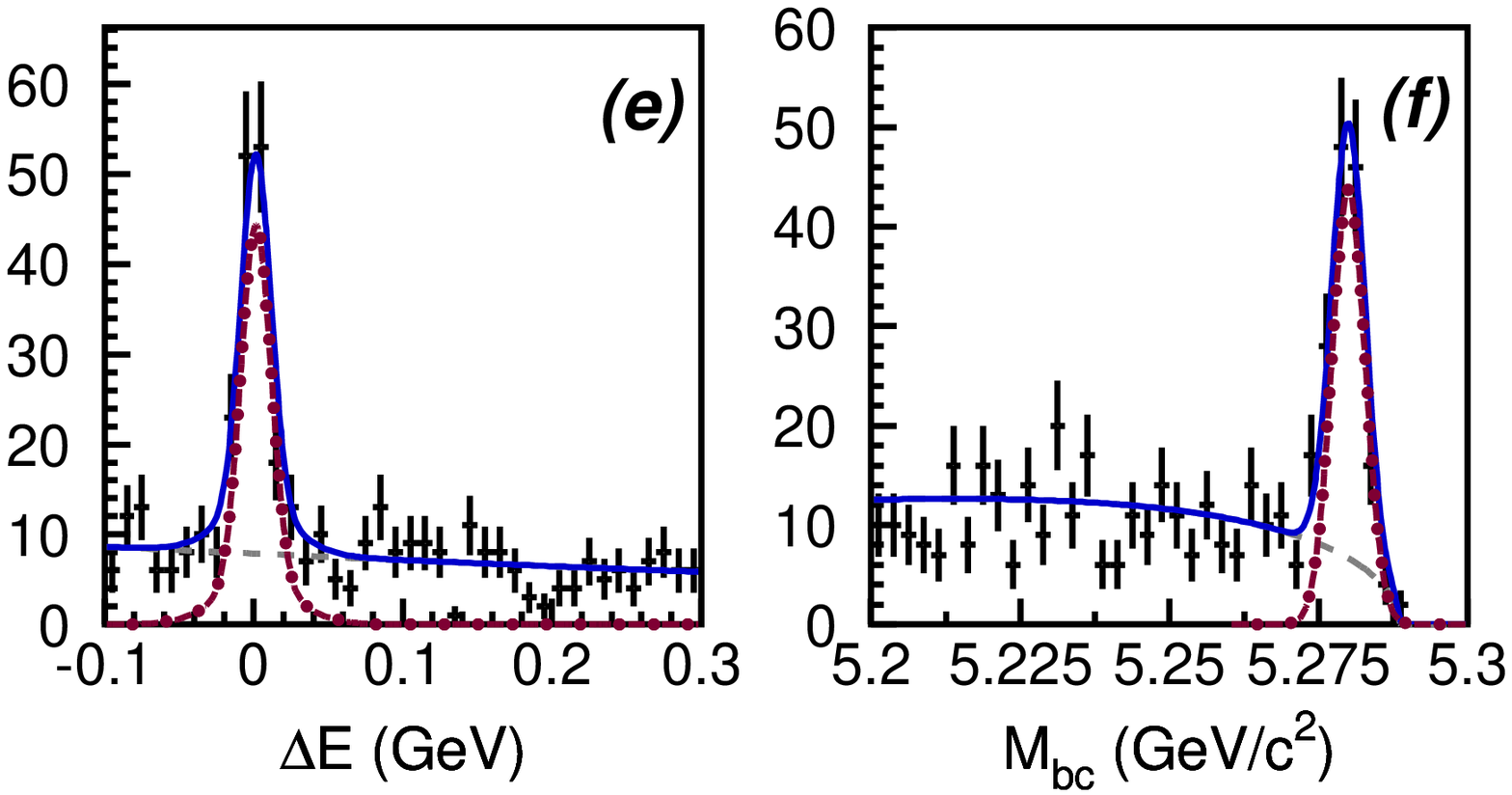, width=3.5in}
\caption{ The $\de$ and $\mb$ distributions for 
(a)(b) $\plg$, (c)(d) $\plpiz$ and (e)(f) $\plpi$ 
modes with the requirement of 
baryon-antibaryon mass $< 2.8$ GeV/$c^2$. The solid curve
represents the fit projection,
which is the sum of signal (dash-dotted peak)
and background (dashed curve) estimations.
}
\label{fg:mergembde}
\end{center}
\end{figure}

Figure~\ref{fg:mergembde} illustrates the fits for the $B$ yields
in a baryon-antibaryon mass region below 2.8 GeV/$c^2$, 
which we refer to
as the threshold-mass-enhanced region. 
The $\mb$ distributions
(with $-0.135$ GeV$<\de< 0.074$ GeV 
for $\pi^0/\gamma$ modes and
$|\de|<$ 0.03 GeV for the $\pi^-$ mode),
 and the $\de$ distributions 
(with $\mb >$ 5.27 GeV/$c^2$)
for the $\plg$, $\plpiz$ and $\plpi$ modes are
shown. 
The solid curves show the projections of the fit results.
%The projections of the fit results are shown  
%by the solid curves. 
%Since there is no clear evidence for the $\ppksz$
%mode, we plot the 90\% confidence level upper limit in this figure.
The $B$ yields are
98 $^{+13}_{-12}$,
56 $^{+11}_{-9}$,
and 129 $^{+14}_{-12}$ 
with
statistical significances of $14.3 $, $9.5$, and $18.9$ standard
deviations 
for the $\plg$, $\plpiz$, and $\plpi$ modes, respectively. 
The significance is defined as $\sqrt{-2{\rm ln}(L_0/L_{\rm max})}$,
where $L_0$ and
$L_{\rm max}$ are the likelihood values returned by the fit with
the signal yield fixed to zero and at its best fit value.
%The measured 
%branching fractions obtained by summing the partial branching fractions 
%in mass bins below 2.8 GeV/$c^2$ are ${\mathcal B}(\bp \to \plg) =
%(2.04^{+0.28}_{-0.26} \pm 0.18) \times 10^{-6}$, 
%${\mathcal B}(\bp \to \plpiz) =
%(1.97^{+0.39}_{-0.32} \pm 0.22) \times 10^{-6}$, 
%and ${\mathcal B}(B^0 \to \plpi) = (2.34 ^{+0.25}_{-0.22} \pm 0.22)  
%\times 10^{-6}$. 
%The 
%significance is defined as $\sqrt{-2 {\rm ln}(L_0/L_{max})}$~\cite{PDG}, 
%where $L_0$ and
%$L_{max}$ denote the likelihood with signal yield fixed at
%zero and at the fitted value, respectively. 
%by the maximum value of the likelihood function
%and the likelihood value with yield fixed at zero~\cite{PDG}.
%The yield for $\ppksz$ is less significant.
%The non-resonant $\ks \pi^+$ component of the 
%$\ppkst$ mode is included in the
%systematic error estimation as described later.

\begin{figure}[htb]
\epsfig{file=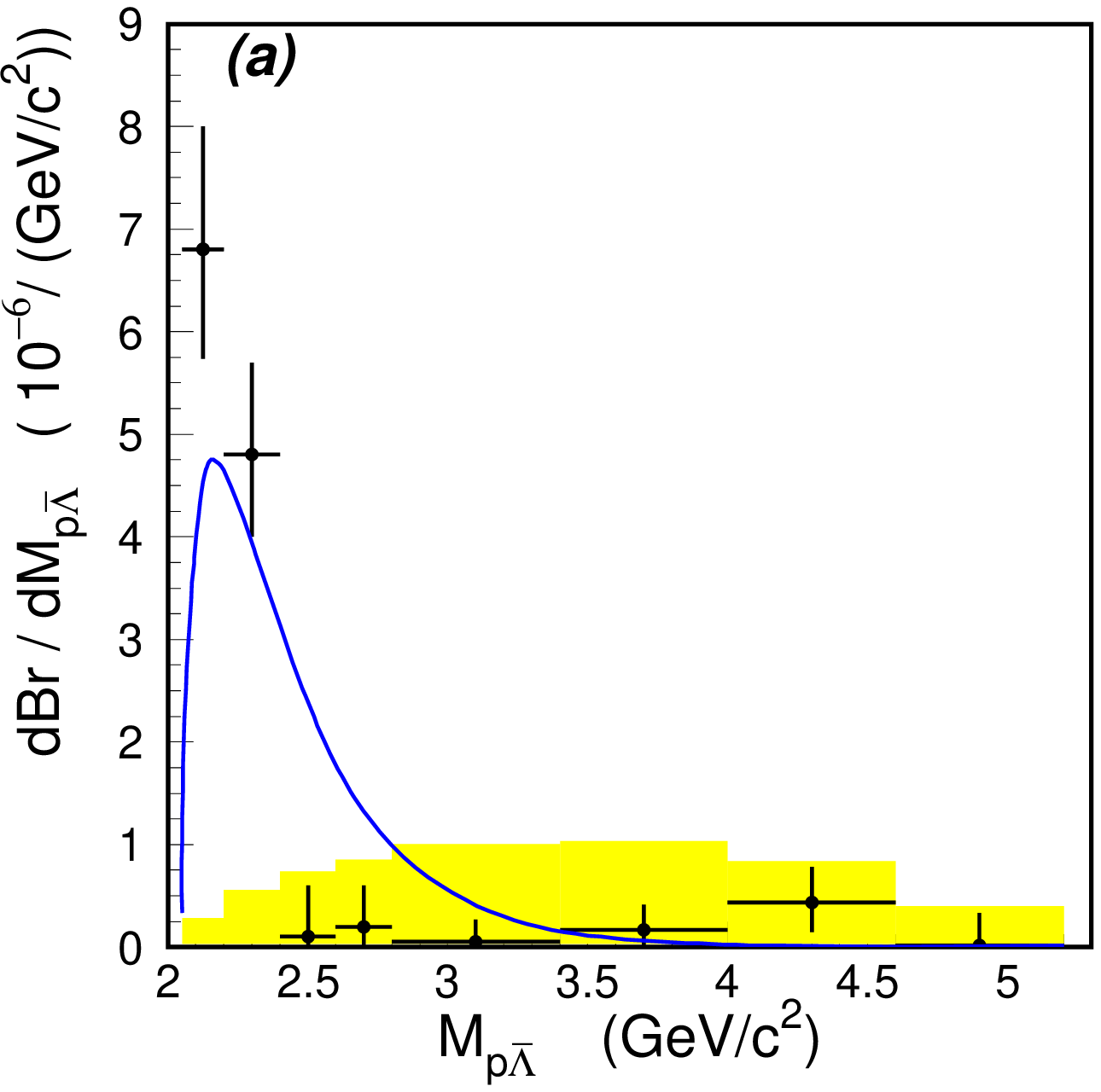, width=2.5in}
\epsfig{file=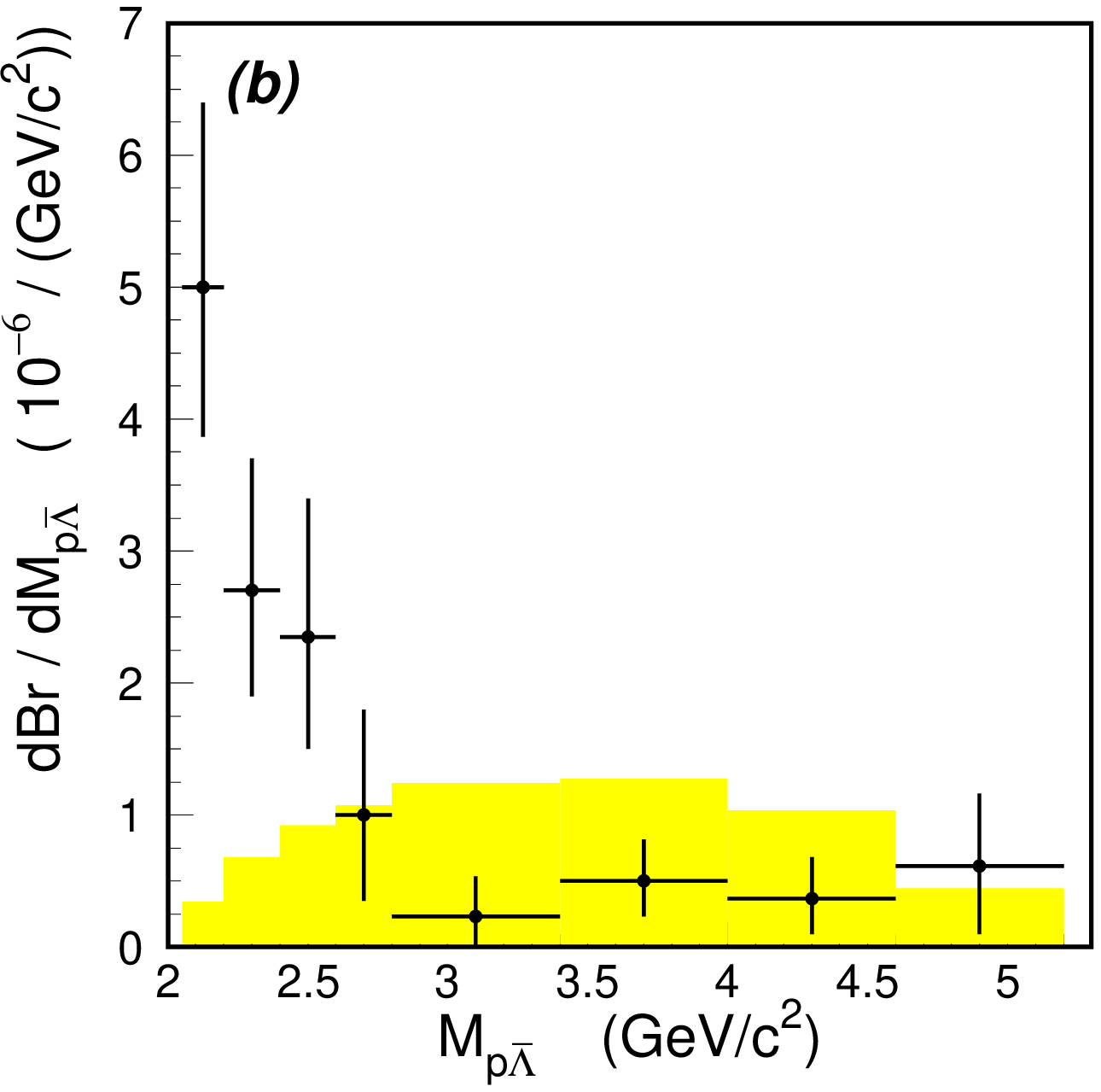, width=2.5in}
\epsfig{file=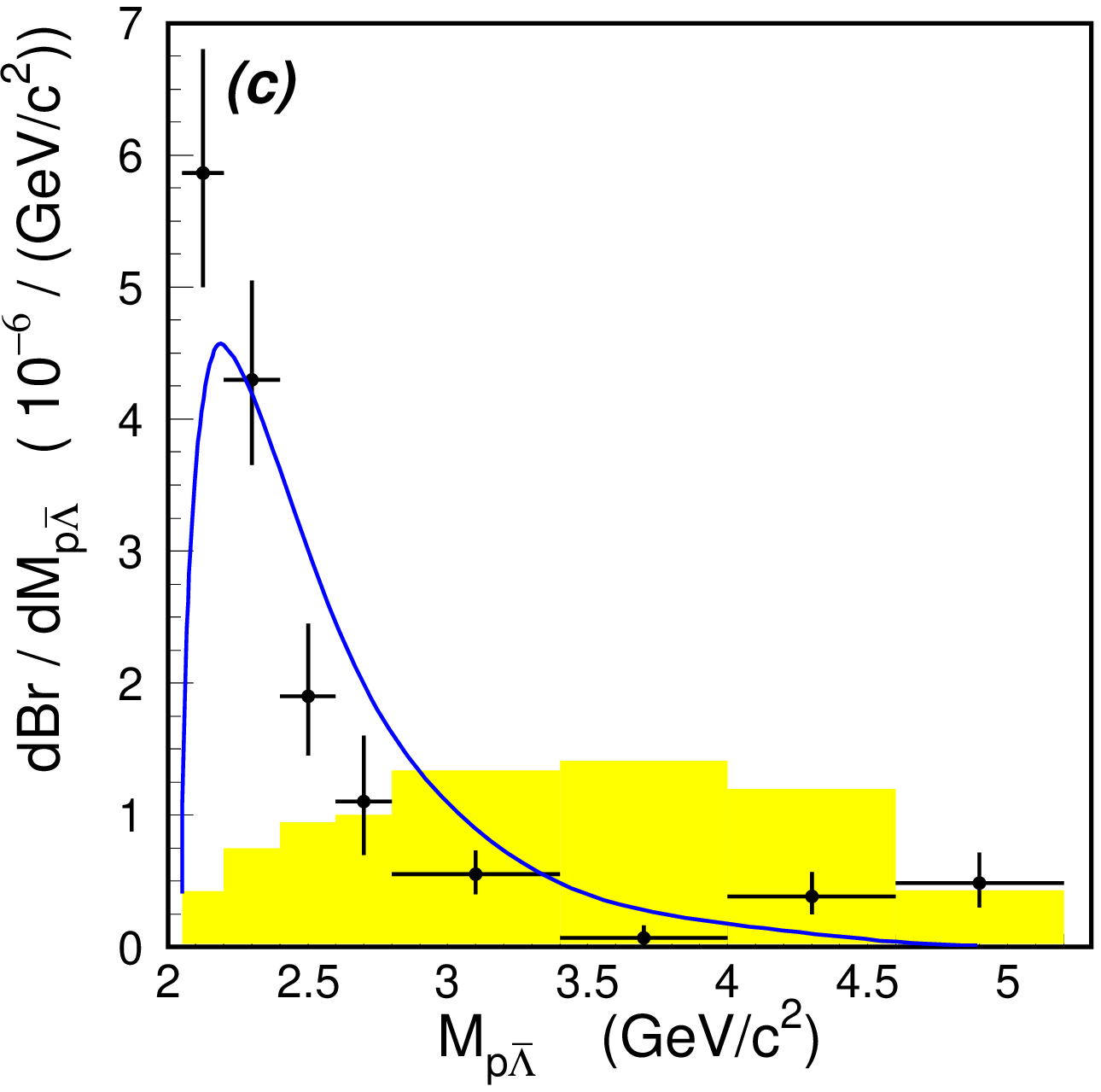, width=2.5in}
\caption{Differential branching fractions for 
%%%divided by the available bin size for
(a) $\plg$, (b) $\plpiz$ and (c) $\plpi$ 
modes %%%in bins
as a function of baryon-antibaryon pair mass. The shaded distribution
shows the expectation from a
phase-space MC simulation. 
The theoretical predicted curves 
from Ref.\ [11] for the $\plg$ mode
and  from
Ref.\ [10] for the $\plpi$ mode  
%~\cite{gengg,HandC} 
are overlaid for comparison. 
The area of the shaded distributions and areas under the theoretical curves 
are scaled to match the measured branching
fractions from data. The uncertainties are statistical only.}
% or theoretical prediction if it is available.
%%to modes with $\pp$ pair.
%The insect shows the $p\bar{p}$ mass distribution for the 
%$J/\psi$ region.

\label{fg:allphase}
\end{figure}

Figure~\ref{fg:allphase} 
shows the 
differential branching fractions of $\bp \to \plg$, $\bp \to \plpiz$ and
$\bz \to \plpi$ as a function of baryon pair mass, 
 where the branching fractions are obtained by correcting
    the fitted $B$ yields for the mass-dependent efficiencies estimated
    from MC simulation for each mode.
Systematic uncertainties %in particle selection
are determined using high-statistics control data samples. 
The tracking efficiency is measured with
fully and partially reconstructed $D^*$ samples.
For proton
identification, we use a  $\Lambda \to p \pi^-$ sample, while for
$K/\pi$ identification we use a $D^{*+} \to D^0\pi^+$,
 $D^0 \to K^-\pi^+$ sample. The average 
efficiency difference for  particle identification  (PID) 
between data and MC has been corrected to obtain the final branching fraction
measurements. The corrections are about 8\%, 8\%, and 14\% for the $\plg$,
$\plpiz$ and $\plpi$ modes, respectively. The uncertainties associated with
the PID corrections are estimated to be 2\% for protons 
and 1\% for charged pions.    
%To be conservative, we take one half of the correction 
%as the systematic uncertainty.   
%the $\mathcal LR$ selection is studied
%with a
%$B^0 \to D^- \pi^+$, $D^- \to \ks \pi^-$ sample
%$\eta \to \gamma\gamma$ and $\eta \to \pi^+\pi^-\pi^0$ samples.
%The $\Lambda$ and $\ks$ reconstruction efficiencies have the same
%uncertainty due to off-IP tracks if the uncertainty of the daughter
%proton identification criterion is not taken into account.
%The uncertainty of $\ks$ reconstruction due to off-IP tracks 
%is determined from a $D^- \to \ks\pi^-$
%sample.
For $\Lambda$ reconstruction, we have an additional 
uncertainty of 2.5\% on the
efficiency for tracks displaced from the interaction point. This is
determined from the
difference between $\Lambda$ proper time distributions for data and MC
simulation. %% This error amounts to 2.5\%.
There is also a 1.2\%
error associated with the $\Lambda$ mass selection and a $0.5\%$ error for the
$\Lambda$ vertex selection. Summing the
errors for $\Lambda$ reconstruction in quadrature, we obtain
a systematic error of 2.8\%.
A 2.2\% uncertainty for the photon detection is determined from radiative
Bhabha events.  % We quote a 2.2\% error for this.
For the $\pi^0$ and
$\eta$ vetoes, we compare the fit results with and without the vetoes;
the difference in the branching fraction is 0.5\%, which is taken
as the associated systematic error. The uncertainty in 
$\pi^0$ reconstruction is studied 
with $ D \to K \pi$ and $D \to K \pi \pi^0$ samples.
The $\mathcal R$ continuum suppression uncertainty is estimated from
$B \to D\pi$, $D \to K^0_S \pi$ control samples, which have 
topologically similar final states.
%$b->c$ control sample
%a $B \to D^0\pi,D^0 \to K\pi$ control sample.
%$\bp \to J/\psi K^+$, $\bz \to J/\psi\ks$, $\bp \to J/\psi\kst$, and
%$\bz \to J/\psi\ksz$
%(with $J/\psi \to \mu^+\mu^-$) control samples.
%Based on these studies, we assign the systematic error and
%we assign a 1\% error for each track, 2\% for each proton identification,
%2\% for each kaon/pion identification, 5\% for $\Lambda$
%ff-IP reconstruction
%and 6\% for the $\mathcal R$ selection.
The determined efficiencies near threshold contribute an error due to 
the binning effect in $M_{\pL}$.
Using the generated MC samples, we vary the bin size to
estimate this effect.
A systematic uncertainty in the fit yield is determined 
by applying different signal/background PDFs 
and by varying
the parameters of the signal and background PDFs by one standard deviation.
The $\plpiz$ mode has a bigger fitting uncertainty due 
to a larger fluctuation
in the lower $\de$ side.    
%We extend the lower bound of $\de$ to $-0.3$ GeV and obtain  
%a consistent result at the 1\% level.
% and is about 4\%. 
%The feed down effect from
%$\ppk$ to $\pppi$ makes the total fit error of the $\pppi$ mode at
%the 5\% level. 
%The MC statistical
%uncertainty %and binning of the baryon pair mass 
%contributes a 2\%
%error in the branching fraction determination. 
The error on the
number of $B\bar{B}$ pairs is 1.3\%, where we
assume  that the branching fractions of $\Upsilon({\rm 4S})$ 
to neutral and charged $B\bar{B}$ pairs are equal. 
%~\cite{Bellenote}
The systematic uncertainties for each 
decay channel are summarized in Table~\ref{systematics}, where
correlated errors are added linearly within each item, and then 
uncorrelated items are combined
in quadrature. The total systematic
uncertainties are 9.0\%, 11.1\% and 9.0\% for
the $\plg$, $\plpiz$ and $\plpi$ modes,
respectively.

\begin{table}[htb]
\caption{Systematic uncertainties of the branching fraction for each decay channel.}
\label{systematics}
\begin{center}
\begin{tabular}{c|ccc}
Source & \ $\plg$ & $\plpiz$ & $\plpi$ 
\\
\hline
Tracking			& \ $4.9\%$ & $4.7\%$ & $5.8\%$
\\
Proton Identification		& \ $4.0\%$ & $4.0\%$ & $4.0\%$
\\
K/$\pi$ Identification		& \ - & - & $1.0\%$ 
\\
BR of $\Lambda\to p \pi^-$ 	& \ $0.8\%$& $0.8\%$& $0.8\%$
\\
$\Lambda$ selection 	 	& \ $2.8\%$& $2.8\%$& $2.8\%$
\\
Photon reconstruction		& \ $2.2\% $ & - & -
\\
$\pi^0$ and $\eta$ veto		& \ $0.5\%$ & - & -
\\
$\pi^0$ reconstruction		& \ - & $4.0\%$ & -
\\
Likelihood Ratio Selection (${\mathcal R}$) & \ $2.5\%$ & $4.0\%$ & $4.0\%$
\\
Modeling and MC statistical error& \ $3.9\%$ & $3.3\%$ & $2.0\%$
\\
Fitting				& \ $2.2\%$ & $5.6\%$ & $1.0\%$
\\
Number of $B\bar{B}$ pairs   	& \ $1.3\%$& $1.3\%$&$1.3\%$
\\
\hline
Total                           & \ $9.0\%$& $11.1\%$&$9.0\%$ 
\\
\end{tabular}
\end{center}
\end{table}

Table~\ref{bins} gives the measured branching fractions
 for different $M_\pL$ mass bins. 
%The differential branching fraction as a function of the 
%baryon pair  mass
%is shown in Fig.~\ref{fg:allphase} and in Table~\ref{bins}.  
%The error bars include the statistical uncertainty from the fit
%and the systematic uncertainty. 
%Here, 
%the efficiency as a function of baryon pair mass for each signal mode is
%determined by MC simulation.
We sum these partial branching fractions to obtain  
${\mathcal B}(\bp \to \plg) = (2.45 ^{+0.44}_{-0.38} \pm 0.22)  
\times 10^{-6}$, 
${\mathcal B}(\bp \to \plpiz) = (3.00 ^{+0.61}_{-0.53} \pm 0.33)  
\times 10^{-6}$, 
and ${\mathcal B}(\bz \to \plpi) = (3.23 ^{+0.33}_{-0.29} \pm 0.29)  
\times 10^{-6}$.
These values are in good agreement with 
our previous measurements~\cite{plg,polar}  
and supersede them.
%They also supersede our previous measurements with better precision.
Note that the results include the
first observation of $\bp \to \plpiz$. 
%The measured branching
%fraction of
%$\bp \to \plpiz$ is much larger than one half of the branching fraction 
%of $\bz \to \plpi$. 
The ratio of ${\cal{B}}(\bp\to\plpiz) /{\cal{B}}(\bz\to\plpi)$ 
is $0.93^{+0.21}_{-0.19} \pm 0.09$, which is larger than the theoretical
prediction of 0.5.
However, one cannot rule out 
the naive factorization picture with current statistics.
The shapes of the near threshold peaks can be
 compared with theoretical predictions~\cite{HandC,gengg}, as shown 
in Fig.~\ref{fg:allphase}.
 This comparison is useful for validating (and possibly modifying)
 theoretical models.

\begin{table}[htb]
\caption{Measured branching
fractions ${\cal B}$($10^{-6}$)
for each $M_\pL$ bin.}
\label{bins}
\begin{center}
\begin{tabular}{c|ccc}
$M_{\pL}$ (GeV/$c^2$)	& \ $\plg$ 	& $\plpiz$ 		& $\plpi$
\\
\hline $ {\rm threshold}-2.2$ & \ $1.02^{+0.18}_{-0.16}$& $0.75^{+0.21}_{-0.17}$& $0.88^{+0.14}_{-0.13}$
\\
 $2.2-2.4$		& \ $0.96^{+0.18}_{-0.16}$& $0.54^{+0.20}_{-0.16}$& $0.86^{+0.15}_{-0.13}$
\\
$2.4-2.6$		& \ $0.02^{+0.10}_{-0.09}$& $0.47^{+0.21}_{-0.17}$& $0.38^{+0.11}_{-0.09}$
\\
$2.6-2.8$		& \ $0.04^{+0.08}_{-0.08}$& $0.20^{+0.16}_{-0.13}$& $0.22^{+0.10}_{-0.08}$
\\
$2.8-3.4$		& \ $0.03^{+0.13}_{-0.11}$& $0.14^{+0.18}_{-0.18}$& $0.33^{+0.11}_{-0.09}$
\\
$3.4-4.0$		& \ $0.10^{+0.15}_{-0.10}$& $0.30^{+0.19}_{-0.16}$& $0.04^{+0.06}_{-0.06}$
\\
$4.0-4.6$		& \ $0.26^{+0.21}_{-0.17}$& $0.22^{+0.19}_{-0.16}$& $0.23^{+0.11}_{-0.10}$
\\
 $4.6-M_{\pL-{\rm lim}}$	& \ $0.01^{+0.19}_{-0.18}$& $0.37^{+0.33}_{-0.31}$& $0.29^{+0.14}_{-0.11}$
\\
\hline
 below 2.8	& \ $2.04^{+0.28}_{-0.26}$& $1.97^{+0.39}_{-0.32}$& $2.34 ^{+0.25}_{-0.22} $
\\
 full region	& \ $2.45 ^{+0.44}_{-0.38}$& $3.00 ^{+0.61}_{-0.53}$& $3.23 ^{+0.33}_{-0.29} $
\\
\end{tabular}
\end{center}
\end{table}

%We also observe the low $\pL$ mass peak near threshold
%for the $\plpiz$ decay as shown in Fig~~\ref{fg:allphase}.
%The curves are 
%theoretical predictions for $\plg$~\cite{gengg} 
%and $\plpi$~\cite{HandC}, which
%represent the low mass peak structures qualitatively.
%, with the events distributed uniformly in phase space.
%%We generate phase space MC samples and determine the average efficiency
%%in bins of baryon pair mass. 
%The widths of the low mass peaks are systematically 
%narrower than
%the theoretical predictions for $\plg$~\cite{gengg} 
%and $\plpi$~\cite{HandC}, 
%as shown in 
%Fig.~\ref{fg:allphase}.
%with proper normalization. 
%This indicates 
%that some modifications
%of theory are needed. 
%other dynamic mechanism 
%should be considered besides the primary $b \to s$ weak decay process. 

We also study the two-body intermediate decays
$\bz \to p {\bar\Sigma}^{*-}$, $\bz \to \Delta^0 \bar{\Lambda}$,
$\bp \to p {\bar\Sigma}^{*0}$, and  $\bp \to \Delta^+ \bar{\Lambda}$, where 
the $\bar\Sigma^{*-,*0}$ and $\Delta^{0,+}$ are reconstructed in the  
$\bar\Sigma^{*-,*0}\to \bar\Lambda\pi^{-,0}$ 
and $\Delta^{0,+} \to p \pi^{0,+}$ channels, respectively. The selection
criteria are 
$1.30$ GeV/$c^2$ $< M_{\bar\Lambda\pi^{-,0}}<1.45$ GeV/$c^2$ and 
$M_{p\pi^{0,+}} < 1.40$ GeV/$c^2$. 
No significant signals are found in these decay chains. 
We observe $34$, $50$, $32$ and $43$ events
in the signal region; the expected number of background events are
$36.9 \pm 1.5$, $51.8 \pm 1.8$, $34.0 \pm 1.3$ and $41.8 \pm 1.2$ for
$\bz \to p {\bar\Sigma}^{*-}$, $\bz \to \Delta^0 \bar{\Lambda}$,
$\bp \to p {\bar\Sigma}^{*0}$, and  $\bp \to \Delta^+ \bar{\Lambda}$, 
respectively. 
We set upper limits 
on the branching fractions at the 90\% confidence
level using the methods described in Refs.~\cite{Gary, Conrad}, where the systematic uncertainty is 
taken into account.
The results are summarized in Table~\ref{br-results}. 
%It is interesting to see
%that the present limits are approaching the $10^{-7}$ level and invite
%new theoretical predictions. 

\begin{figure}[htb]

\epsfig{file=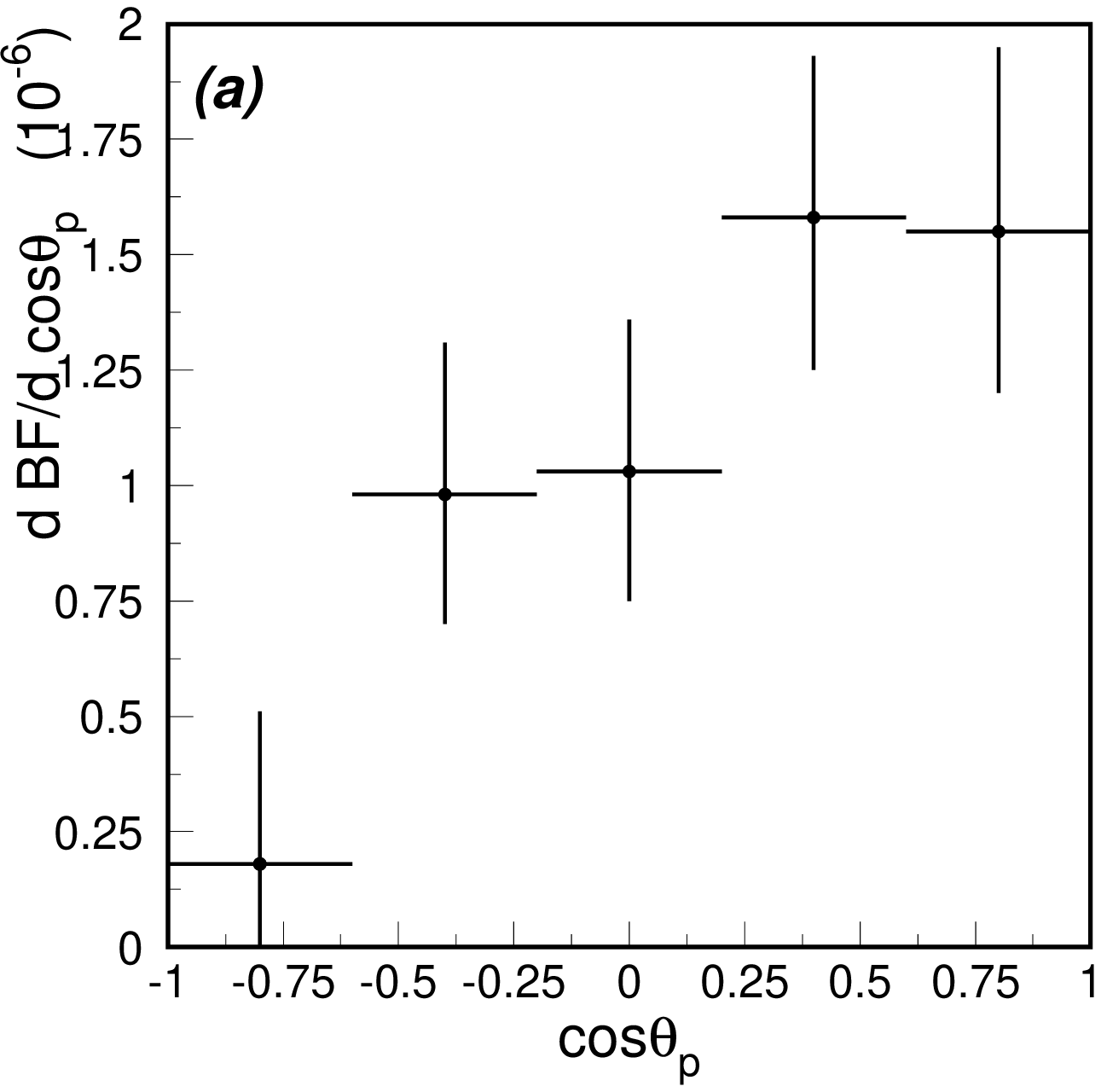, width=2.5in}
\epsfig{file=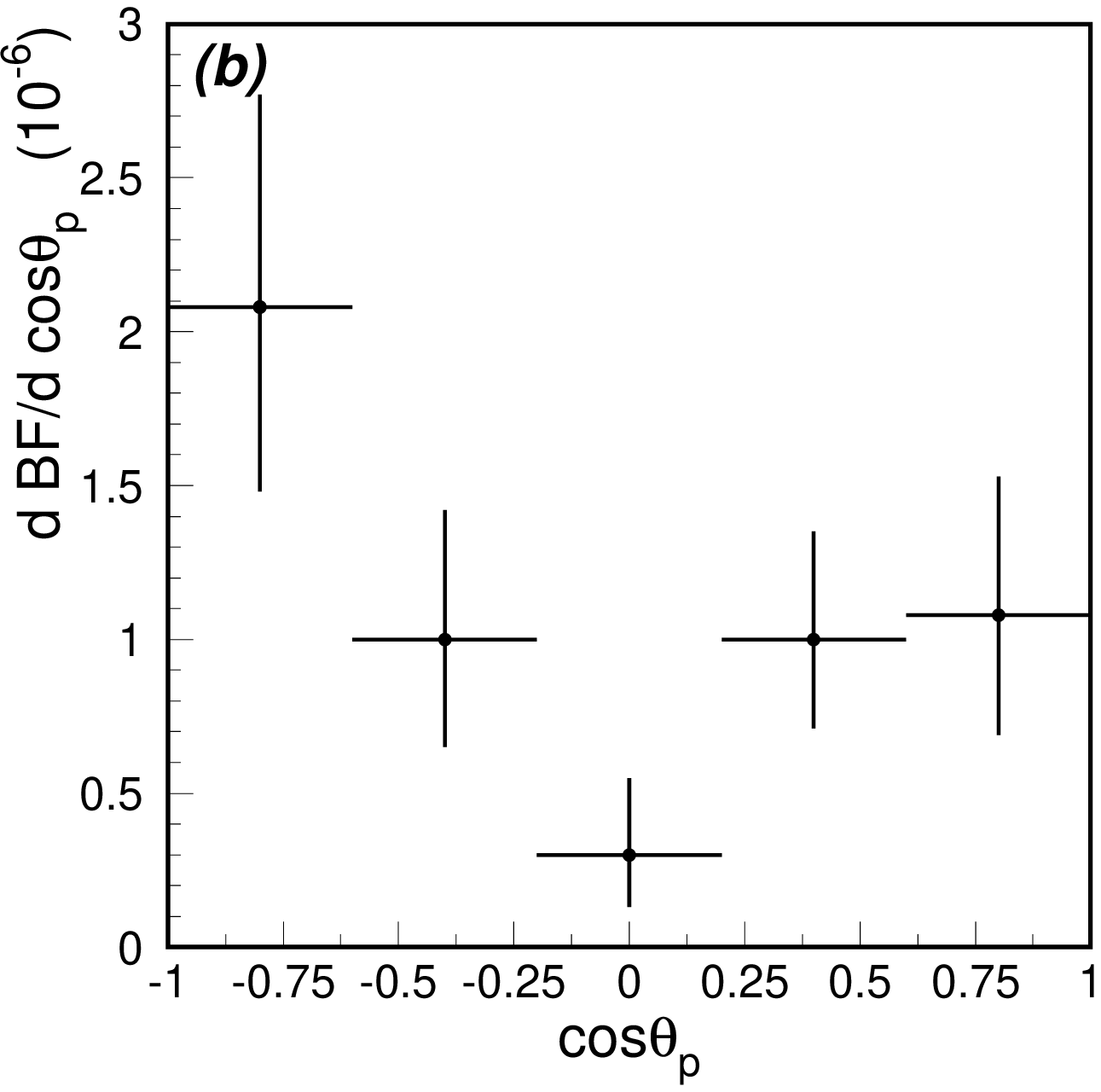, width=2.5in}
\epsfig{file=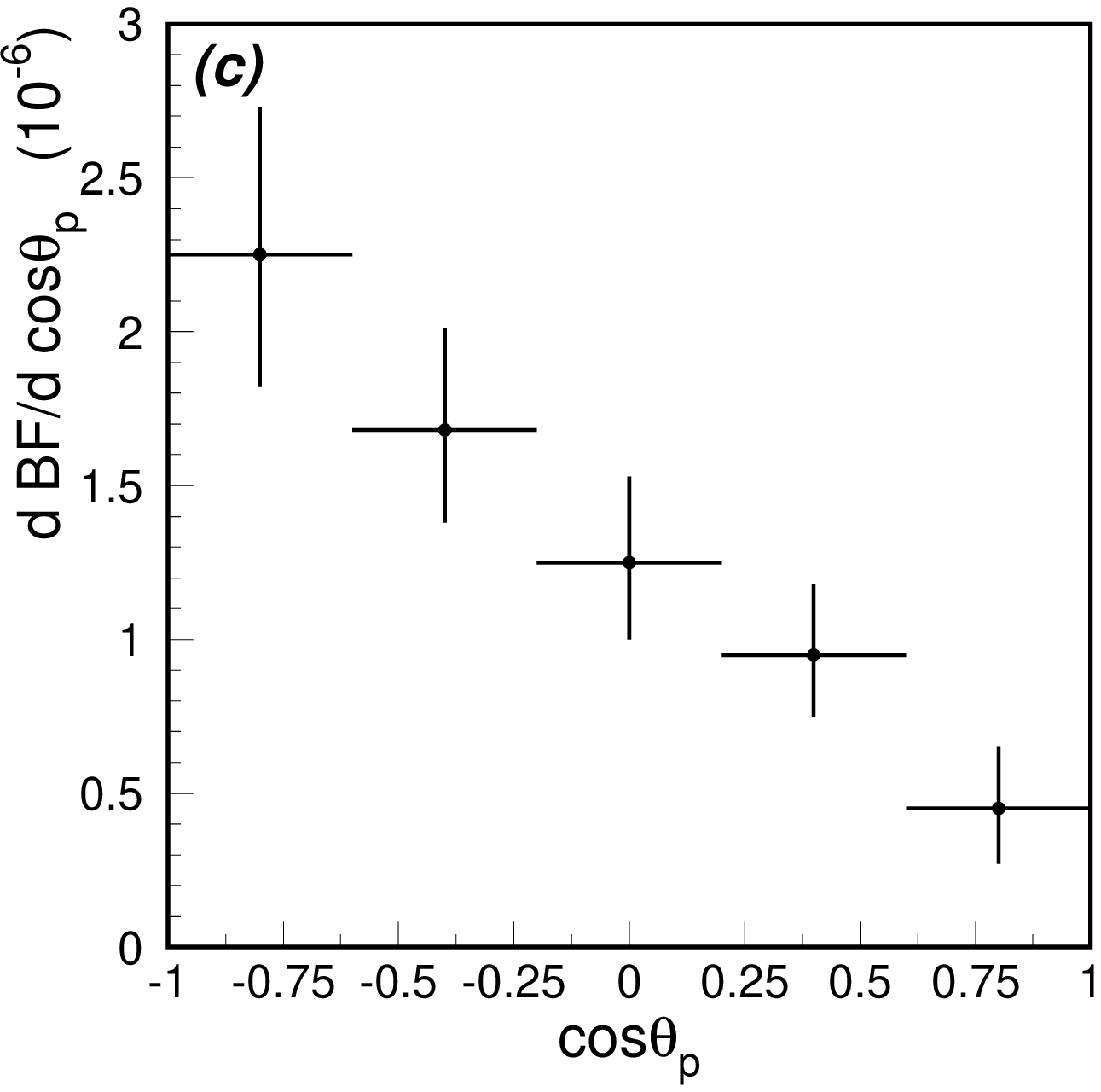, width=2.5in}
\caption{Differential branching fractions {\it vs.} $\cos\theta_p$
% with baryon-antibaryon mass $< 2.8$ GeV/$c^2$ 
for (a) $\plg$, (b) $\plpiz$ and and (c) $\plpi$ 
modes in the region near 
threshold (baryon-antibaryon mass $< 2.8$ GeV/$c^2$). The uncertainties are
statistical only.}
%%to modes with $\pp$ pair.
%The insect shows the $p\bar{p}$ mass distribution for the 
%$J/\psi$ region.
\label{fg:thetap}

\end{figure}

In the low mass region below 2.8 GeV/$c^2$,
we study the proton angular distribution of the baryon-antibaryon pair system.
%in its helicity 
%frame. 
The angle $\theta_p$ is defined as the
angle between the proton direction and the meson (photon) direction in the 
baryon-antibaryon pair rest frame.  
Figure~\ref{fg:thetap} shows the differential
branching fractions as a function of $\cos\theta_p$. 
%The error bars
%include the statistical uncertainty from the fit
%and the systematic uncertainty. 
We define the angular asymmetry as $A_{\theta} = {
{Br_+ - Br_-}\over
{Br_+ + Br_-}}$, where $Br_+$ and
$Br_-$
stand for the measured branching fractions with $\cos\theta_p > 0$ and
 $\cos\theta_p < 0$, respectively. The angular asymmetries are determined to be
$0.29 \pm 0.14({\rm stat.}) \pm 0.03({\rm syst.})$, 
$-0.16 \pm 0.18({\rm stat.}) \pm0.03({\rm syst.})$,
and $-0.41 \pm 0.11({\rm stat.}) \pm 0.03({\rm syst.})$ for the $\plg$, $\plpiz$, and $\plpi$ modes, 
respectively. 
%The above uncertainty is predominantly due to statistical error
%and the systematic error is negligible.
A systematic error, $\sim 0.03$, is determined 
by studying low momentum $\Lambda$ reconstruction in 
different angular regions,  
and by checking the $\bp \to J/\psi K^+$ ($J/\psi \to
\mu^+\mu^-$) sample and the continuum background of $\bp \to \ppk$ 
where a null asymmetry is expected. 

%One important contribution about
%the agreement between data and MC for slow $\Lambda$ simulation
%is checked by flight distance   
%From the asymmetric results, we find that the threshold enhancement peak
Since $A_{\theta}$ is not consistent with zero for $\bz \to \plpi$, 
the peak near threshold
cannot be described by a single resonant state~\cite{MSuzuki}. 
The opposite slopes in the distributions for the $\plg$ and $\plpi$ modes 
indicate that the $\plg$ decay agrees well with the short-distance 
$b \to s \gamma$
picture while the $\plpi$ mode 
disagrees with the short-distance $b \to s g$  description, where $g$ stands 
for a hard gluon. 
%Usually the $\bar\Lambda$ from the $\plpi$ decay mode travels
%parallel to a faster proton in the $B$ meson rest frame.
The low mass peaking structure in $M_\pL$ implies that $p$ and $\bar\Lambda$ 
are moving in parallel in the $B$ meson rest frame. 
%One can tell 
%which baryon is moving faster by boosting to the dibaryon rest
%frame and check the plot show in Fig.~\ref{fg:thetap}. We then found that 
%$\bar\Lambda$ moves slower for the $\plpi$ case.
One can look for correlations using the angular distributions in
Fig.~\ref{fg:thetap}.  The negative slope for the $\plpi$ mode in
Fig.~\ref{fg:thetap}c implies that the proton moves faster 
and the $\bar{\Lambda}$ moves slower.
In other words, the $s$ quark from $b$ decay is not as energetic as expected.
Disagreement between data and the short-distance description
 has already been found in the decay $\bp \to \ppk$~\cite{polar}.  
One possible explanation is the contribution of long-distance effects.
%Our previous publication~\cite{polar} for $\bz \to \plpi$
%does not reveal this because of
%low statistics. 
%This interesting feature should be studied with much larger
% data sample in separate $\mpl$ regions.   
%The systematic error is included in the above uncertainty.

\begin{figure}[b!]

\epsfig{file=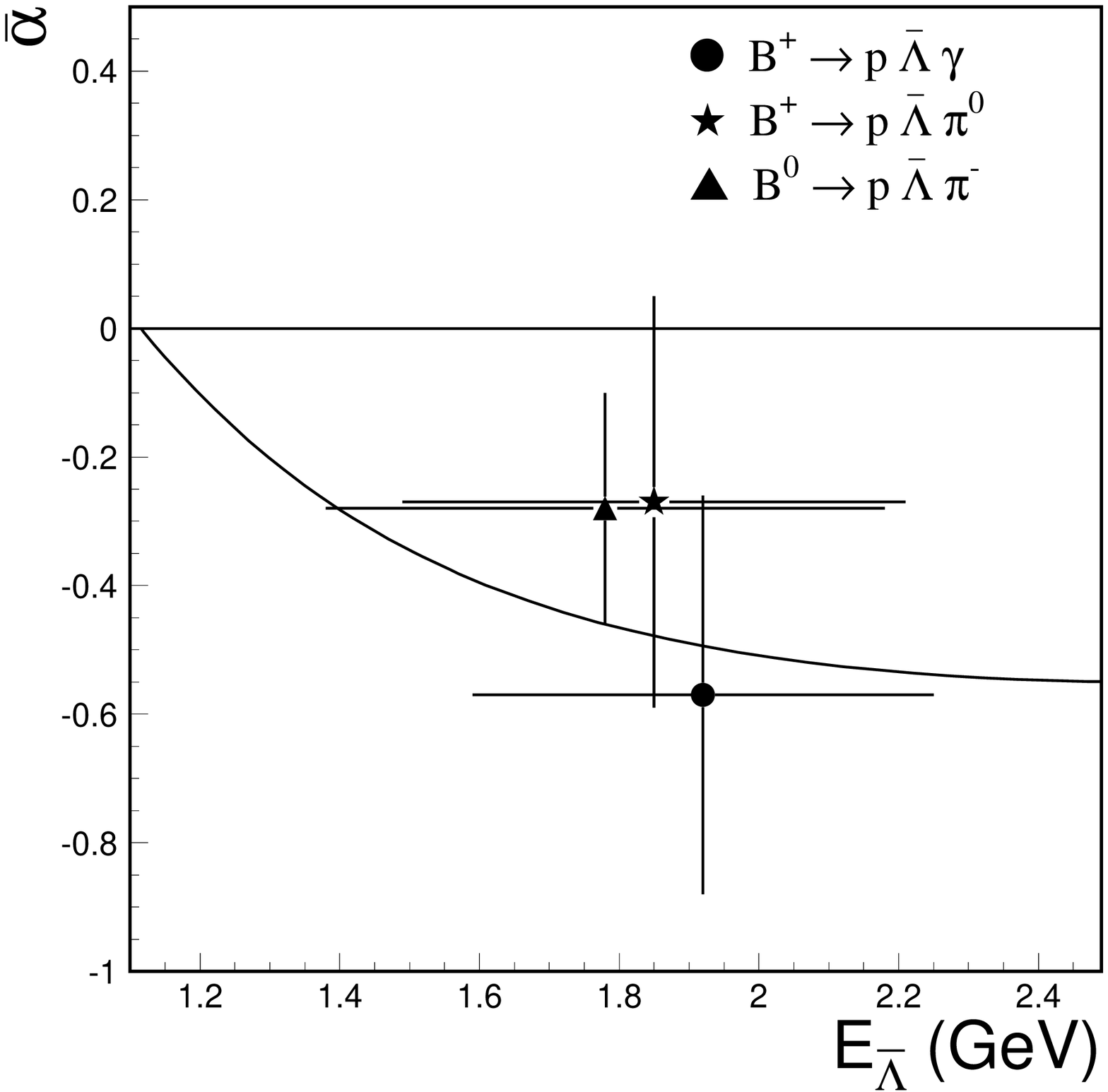, width=3in}
\caption{Anisotropy parameter $\bar\alpha$ vs. E$_\Lambda$ in the $B$ rest frame
% with baryon-antibaryon mass $< 2.8$ GeV/$c^2$ 
for $\plg$, $\plpiz$ and $\plpi$ 
modes. The energy spread for each decay mode is represented 
by the horizontal error bar.
The theoretical prediction by Ref. [12] is shown as a solid curve.}
%%to modes with $\pp$ pair.
%The insect shows the $p\bar{p}$ mass distribution for the 
%$J/\psi$ region.
\label{fg:effectalpha}

\end{figure}

Another interesting feature of $B$ decays with a 
$\Lambda$ in the final state
is the possibility of
 using the $\Lambda$ as a helicity analyzer of the $s$ quark 
in order to check the 
left-handedness of $b \to s$ weak decays. 
We modify the unbinned likelihood fit in order to simultaneously estimate 
the anisotropy parameter
of the secondary proton from $\Lambda$ decays. 
The parameterization is
$1 + \bar\alpha \cos\theta$, where $\bar\alpha$ is the parameter and
$\theta$ is the
angle between the secondary proton momentum and the direction opposite
to the $B$ momentum in the $\Lambda$ rest frame. Note that the anisotropy 
parameter $\bar\alpha$
is identical for both $\Lambda$ and $\bar\Lambda$.
The measured values are 
%$-0.71 \pm 0.34$, $0.21 \pm 0.32$ and 
%$ -0.27 \pm 0.20$
$-0.57 \pm 0.33({\rm stat.}) \pm 0.10({\rm syst.})$,
$-0.27 \pm 0.33({\rm stat.}) \pm 0.10({\rm syst.})$,
and $ -0.28 \pm 0.21({\rm stat.}) \pm 0.10({\rm syst.})$
for the $\plg$, $\plpiz$ and $\plpi$ modes, 
respectively. The average $\Lambda$ energies in the $B$ rest frame are
determined to be 
$1.92$ GeV, $1.85$ GeV, and $1.78$ GeV with standard deviations
 of $0.33$ GeV, $0.36$ GeV, and $0.40$ GeV for the $\plg$, $\plpiz$ and $\plpi$ modes, respectively.
%The energy uncertainty stands for the variance of the energy spread.
Figure~\ref{fg:effectalpha} shows the measured anisotropy parameters for
different decay modes and compares the results with 
the prediction of the Standard Model~\cite{Suzuki} as a 
function of $\Lambda$ energy. They are
consistent within errors. The value of $\bar\alpha$ obtained  
for the $\plpi$ mode
also agrees well with the theoretical prediction in Ref.~\cite{HandC}.
%for a left-handed quark picture in 
%weak decays.
%They agree well with 
%each other.
% which indicates that the non-Standard Model contribution in
%$b \to s$ is not big. 
The systematic uncertainty in $\bar\alpha$ 
is included in the plot and is about $0.10$. This 
is estimated by varying various selection cuts;
the dominant effect is
the efficiency change near the $\cos\theta \sim 1$ region, 
where the detection efficiency
for slow pions is rapidly changing.    

We also measure the charge asymmetry
as $A_{CP}$= $(N_{b} - N_{\bar{b}})/ (N_{b} + N_{\bar{b}})$ for these modes,
where $b$ stands for the quark flavor of the $B$ meson. The results
are included in Table~\ref{br-results}.
The measured charge asymmetries are consistent with zero within 
their statistical uncertainties. The systematic uncertainty
is assigned by the measured asymmetry of the background events
in the candidate region.

%The asymmetry of the distribution indicates that the fragmentation
%picture is favored. 
%%since a symmetric distribution is 
%%preferred for the gluonic picture. 
%Antiprotons are emitted along
%the $K^+$ direction most of the time, which can be explained by 
%a parent $\bar{b} \to \bar{s}$ penguin transition followed by
%$\bar{s} u$ fragmentation into the final state as shown in Fig.~\ref{fg:feyn}. 
%The energetic $\bar{s}$ quark picks up the $u$ quark from a
%$u\bar{u}$ pair in vacuum and the remaining $\bar{u}$ quark then
%drags a $\bar{u}\bar{d}$ diquark out of vacuum. 
%This simple picture can describe the
%$\bar{p}-K^+$ angular correlation. The spectator $u$ quark and leftover
%$u d$ diquark form an proton.     

\begin{table}[htb]
\caption{Summary of the measured 
results for $\bp \to \plg$, $\plpiz$ and $\bz \to \plpi$. 
Y is the fitted signal or upper limit 
at 90\% confidence, $\sigma$ is the statistical significance, 
{$\cal{B}$} is the branching fraction, 
$A_{\theta}$ is the angular asymmetry and
$A_{CP}$ is the charge asymmetry.}
\label{br-results}
\begin{center}
\begin{tabular}{c|ccccc}
Mode & \ Y & $\sigma$ & {$\cal B$} ($10^{-6}$)  & $A_{\theta}$  & $A_{CP}$ \\
\hline $\bp\to\plg$ & \ $114^{+18}_{-16}$ & 14.5 & $2.45^{+0.44}_{-0.38}\pm 0.22$ & 
$0.29 \pm 0.14 \pm 0.03$ & $0.17\pm 0.16 \pm0.05$
\\
\hline $\bp\to\plpiz$ & \ $89^{+19}_{-17}$ & 10.2 & $3.00^{+0.61}_{-0.53}\pm 0.33$ &
$-0.16 \pm 0.18 \pm 0.03$ & $0.01 \pm 0.17 \pm 0.04$ 
\\
       ~~~~~~~~~~$\bp \to p {\bar\Sigma}^{*0}$ & \ $<11.3$& - & $<0.47$ & - & -
\\
       ~~~~~~~~~~$\bp \to \Delta^+ \bar{\Lambda}$ & \ $<15.9$ & - &$<0.82$ & - & -
\\
\hline $\bz\to\plpi$ & \ $178^{+18}_{-16}$ & 20.0 & $3.23^{+0.33}_{-0.29}\pm 0.29$ &
$-0.41 \pm 0.11 \pm 0.03$ &  $-0.02 \pm 0.10 \pm 0.03$ 
\\
       ~~~~~~~~~~$\bz \to p {\bar\Sigma}^{*-}$ & \ $<10.9$ & -& $<0.26$ & - & -
\\
       ~~~~~~~~~$\bz \to \Delta^0 \bar{\Lambda}$ & \ $<15.9$ & -&$<0.93$ & - & -
\\
\end{tabular}
\end{center}
\end{table}

In summary, using 449 $ \times 10^6 B\bar{B}$ events, we measure the
mass and angular distributions of the baryon-antibaryon pair
system near threshold for the $\plg$,  $\plpiz$ and $\plpi$  
baryonic $B$ decay modes. 
We report the observation of $\bp\to\plpiz$ with
a branching fraction 
$(3.00^{+0.61}_{-0.53}\pm 0.33) \times 10^{-6}$ and a low $\pL$ 
mass peak near threshold. 
The measured branching fractions for $\bp \to \plg$ and $\bz \to \plpi$
are in good agreement with our previous measurements~\cite{plg,polar}.
%The narrow
%width of the threshold enhancement of the  $\plg$/$\plpi$ modes
%cannot be described by available theoretical models.
The different proton polar angular distributions for the $\plg$ and
$\plpi$ modes indicate a difference between $b \to s \gamma$ 
and $b \to s g$ decays. The anisotropy 
parameters $\bar\alpha$ from $\Lambda$ decays
agree with theoretical predictions within errors.
%It plays an important role  
%for the threshold enhancement effect. 
%It will be interesting to compare
%with other observed decay modes (such as $\pppi$ and $\llk$) to learn
% more about the underlying dynamics. 
%We need more observations to understand the underlying dynamics.
We also search for intermediate two-body decays and find no significant
signals. We set upper limits on their 
branching fractions at the 90\% confidence
level. Some 
suppression factors~\cite{chengsuppress} 
for the charmless baryonic two-body decays should be 
considered under the present theoretical framework
and understanding the mechanism of the threshold enhancement 
might be the key to determine the two-body decay rates .

%\section*{Acknowledgments}
We thank the KEKB group for the excellent operation of the
accelerator, the KEK cryogenics group for the efficient
operation of the solenoid, and the KEK computer group and
the National Institute of Informatics for valuable computing
and Super-SINET network support. We acknowledge support from
the Ministry of Education, Culture, Sports, Science, and
Technology of Japan and the Japan Society for the Promotion
of Science; the Australian Research Council and the
Australian Department of Education, Science and Training;
the National Science Foundation of China and the Knowledge
Innovation Program of the Chinese Academy of Sciences under
contract No.~10575109 and IHEP-U-503; the Department of
Science and Technology of India; 
the BK21 program of the Ministry of Education of Korea, 
the CHEP SRC program and Basic Research program 
(grant No.~R01-2005-000-10089-0) of the Korea Science and
Engineering Foundation, and the Pure Basic Research Group 
program of the Korea Research Foundation; 
the Polish State Committee for Scientific Research; 
%-> remove for now: under contract No.~2P03B 01324; 
the Ministry of Education and Science of the Russian
Federation and the Russian Federal Agency for Atomic Energy;
the Slovenian Research Agency;  the Swiss
National Science Foundation; the National Science Council
and the Ministry of Education of Taiwan; and the U.S.\
Department of Energy.


\begin{thebibliography}{99}
\bibitem{conjugate} {Throughout this report,
inclusion of charge conjugate mode is always implied unless
otherwise stated.}
\end{thebibliography}
\end{document}